
\magnification\magstep1
\baselineskip=16pt
\language=0
\def\no{\noindent}

\def\b{\beta}
\def\g{\gamma}
\def\l{\lambda}
\def\n{\nu}
\def\s{\sigma}
\def\p{\psi}
\def\P{\bar{\psi}}
\def\o{\omega}
\def\L{\Lambda}
\centerline{\bf Competition of Phonons and Magnetic Interaction in
One-Dimensional}
\centerline{\bf Fermion Systems}
\vskip 0.2in
\centerline{D. Braak, F. Sievers and K. Ziegler}
\vskip 0.3in
\centerline{Institut f\"ur Theorie der Kondensierten Materie,
Universit\"at Karlsruhe,}
\centerline{ Physikhochhaus, D-76128 Karlsruhe, Germany}
\centerline{(\today)}
\vskip 1.2in
\no
Abstract:\par
\no
We consider a system of spin-dependent fermions on a one-dimensional lattice
which is coupled to phonons. The phonons create either
a Peierls instability by
breaking the translational invariance or create long range correlations.
On the other hand, there is a spin-dependent Hubbard-like interaction which
competes with the effective fermion-fermion interaction induced by the
phonons. We investigate the competition of both interactions using a
renormalization group calculation for weak interaction.
The renormalization group transformation creates a third interaction.
It turns out that non-interacting fermions represent a semi-stable fixed
point of the fermion system.
\vfill
\eject
\no
One-dimensional electronic systems are sensitive to the interaction with
dynamical fluctuations (e.g. phonons). There is the well-kown Peierls
instability [1] which corresponds to a spontaneous breaking of the
translational invariance along the one-dimensional chain through
dimerization. This effect plays a crucial role
in one-dimensional conductors like polyacetylene [2] because it opens a gap
in the electronic spectrum. This phenomenon has been studied extensively
using classical solutions for the dimerization field [3] or using
the renormalization group method [4]. The latter indicates either
long-range correlations (this case was discussed in [5] by means of
a bosonization method) or a
dimerized state through a renormalization group flow towards strong
electron-phonon coupling [6].
An important question for a better understanding
of the electronic properties is the stability of the phonon effect. There
is some indication that a gap will be destroyed by static disorder
[7,8,9]. Disorder can be due to, e.g., doping of polyacetylene
[10]. Another important effect
in one-dimensional electronic systems is a spin-dependent interaction
between the fermions. This also tends to destroy the phonon effect. In the
following we will discuss the competition of the fermion-phonon and a
Hubbard-like interaction. For this
purpose it is convenient to consider an effective fermion-fermion
interaction induced by the fermion-phonon coupling with a coupling
constant $\l$. This interaction acts between fermions sitting on nearest
neighbor sites of the lattice. Therefore, it is useful
to describe this by introducing a sublattice representation of the
fermions. The spin-dependent interaction, on the other hand, is local with
respect to the lattice, characterized by another coupling constant.
Assuming weakly interacting fermions we can apply perturbation theory.

The partition function for the dynamical model of electrons with spin
$\s=1,2,...,n$ on a chain with $2N$ sites reads
$$
{\cal{Z}}=\int {\cal{D}}[\P^\s_a (r,\tau)\,\p^\s_a (r,\tau)]  \exp(-S)
\eqno (1)
$$
with action
$$
S=\int_0^{\b}d\tau\sum_{\s=1}^n\sum_{r=1}^N\sum_{a =1}^2\Big\{
\P _a^\s(r,\tau)\, [i\partial_\tau +\mu ]\, \p _a^\s(r,\tau)
$$
$$
+ \sum_{r'}^N[\P^\s_1(r,\tau)\,d_{rr'}\p^\s_2(r',\tau)+\P^\s_2(r,\tau)
\,d^T_{rr'}\p^\s_1(r',\tau)] +
\l \sum_{b}\P ^\s _a (r,\tau)\,\p ^\s_{b} (r,\tau)\,\P ^\s_{b} (r,\tau)
\,\p^\s_{a}(r,\tau)
$$
$$
+ \g \sum_{\s'=1}^n\P^\s_a (r,\tau)\,\p^\s_a (r,\tau)\,
\P^{\s'}_a(r,\tau)\,\p^{\s'}_a(r,\tau)\Big\}.\eqno (2)
$$
The $\P ,\p$ are real Grassmann variables with sublattice indices
$a,b$ and electron spin indices $\s,\s'$. $d_{r,r'}$ is the
lattice difference operator of the sub-chain with $N-$sites and $\mu$ the
chemical potential.

\no
The coupling $\g$ belongs to a Hubbard-like interaction in $S$. $\l$ is
the coupling constant which belongs to the continuum approximation of
the electron-phonon interaction ($\l<0$). In general, we consider both
coupling constants in the positive and negative regime.

\no
At $T = 0$ and in large scale approximation, the harmonic part
of the action (2), which is the inverse electron propagator of the model,
becomes in momentum/frequency-space
$$
H_0 = \pmatrix{-\o +\mu & ik \cr
-ik & -\o +\mu}
\delta_{\s,\s'}\eqno(3)
$$
where the $2\times2$ matrix structure is due to the sublattice structure.
We confine our consideration to the half-filled case which
corresponds to $\mu =0$.
Introducing an ultraviolet cut-off $\L$ for both the $\o$- and
$k$-integration, we derive the following RG flow equations to
one-loop order by lowering the cut-off to $\L' =\L e^{-l}$[11]:
$$
\eqalignno{
{d\l\over{dl}} &= -\l ^2-(n-1)\,(\n ^2 +\g ^2)
&(4) \cr
{d\g\over{dl}} &= -2\,\g\l-2\,(n-2)\,\g\n
&(5) \cr
{d\n\over{dl}} &= -2\,\l\n-(n-2)\,(\n ^2 + \g ^2)
&(6) \cr
}
$$
It is important to notice that the renormalization procedure generates a new
coupling $\n $ which is conjugate to the operator
$$
O_\n = \P^\s_a(r,\tau)\p^\s_a(r,\tau)\P^{\s'}_b(r,\tau)\p^{\s'}_b(r,\tau)
\eqno(7)
$$
with $a\neq b $ and $\s \neq \s' $. In calculating the resulting
RG flows, the initial value of $\n $ equals zero.
For a qualitative understanding it is useful to consider two special cases
for the initial values $\l_0$, $\g_0$ of the coupling constants.

\no
\item{(i)} $\g_0=0$: $\g$ remains zero according to (5). The flow of the
other coupling constants is
$$
\eqalignno{
{d\l\over{dl}} &= -\l ^2-(n-1)\,\n ^2
&(4a) \cr
{d\n\over{dl}} &= -2\,\l\n-(n-2)\,\n ^2
&(6a) \cr
}
$$

\noindent
According to (6a) $\n$ remains also zero because of the initial value
$\n_0=0$. Thus this case is reduced to the fermion-phonon system.
$|\l|$ increases (decreases) if the initial value is $\l_0<0$ ($\l_0>0$),
respectively. The flow to strong coupling for $\l_0<0$ reflects the Peierls
instability or long-range correlations due to phonons.

It is interesting
that even with $\n_0\ne0$ the renormalization is effectively a
fermion-phonon system. For instance, for $n=2$ (4a) and (6a) can be combined
to one equation for $\l+\n$:
$$
{d(\l+\n)\over{dl}} =-(\l+\n)^2. \eqno (8)
$$
For $n=3$ these equations lead for the effective coupling $\l-\n$ to:
$$
{d(\l-\n)\over{dl}} =-(\l-\n)^2. \eqno (9)
$$
\item{(ii)} $\l_0=0$: the renormalization of $\l$ (4) flows always to strong
negative coupling $\l$ for non-zero $\g_0$.
Thus the Hubbard-like interaction creates an effective fermion-phonon
interaction. It does not renormalize itself in one-loop approximation.

\noindent
The flow of the coupling constants for the initial value $\l_0=0.1$ is
plotted in Figs.1,2. If $\g_0$ is sufficiently weak we approach the
non-interacting fermion fixed point. However, for $|\g_0|$ larger than a
certain threshold (the latter depends on $\l_0$) a strong coupling behavior
sets in. In this case all three coupling constants are driven to strong
interaction. This behavior indicates a transition from non-interacting
fermion to a system of strongly interacting fermions for a critical value
of $g_0$. On the other hand, for $\l_0<0$ the coupling constants are
always driven away from the non-interacting fixed point. An exeptional
case is $\g_0=0$ and $\l_0<0$, where we obtain a system which is governed
by a strong fermion-phonon interaction alone.

In conclusion, the renormalization group flow depends strongly on the
initial values of the two coupling constants $\l_0$ and $\g_0$. We discover
a semi-unstable fixed point $\l=\g=\n=0$ which separates non-interacting
fermions from the flow to strong coupling behavior. The former destructs
entirely the effect of the phonons whereas the latter reflects the
creation of a dimerized state or a state with long-range correlation
for $\l\ll 0$. However, this behavior is accompanied by a strong coupling
$\n$ and $\g$ if $\g_0\ne0$. I.e., the phonons
are not dominating the renormalization group behavior alone even in the
strong coupling regime. Consequently, in general it is not sufficient to
consider only
the fermion-phonon interaction in a one-dimensional conductor but we have
to take into account at least a spin-dependent interaction no matter how
small $\g_0\ne0$ is.

\bigskip
\no
This work was partly supported by the Deutsche Forschungsgemeinschaft
through the Sonderforschungsbereich 195.

\vfill
\eject
{\bf References}

[1] R.E.~Peierls, {\sl Quantum theory of Solids}, Clarendon Press Oxford
(1955)

[2] several articles in: {\sl Physics in One Dimension}, J.~Bernasconi and

$\ \ \ $T.~Schneider (eds.), Springer Verlag Berlin (1980)

[3] W.P.~Su, J.R.~Schrieffer and A.J.~Heeger, Phys.Rev.B22, 2099 (1980)

[4] J.~Solyom, Adv.Phys.28, 201 (1979)

[5] B.~Sakita and K.~Shizuya, Phys.Rev.B42, 5586 (1990)

[6] G.T.~Zimanyi, S.A.~Kivelson and A.J.~Luther, Phys.Rev.Lett.60, 2089
(1988)

$\ \ \ $S.A.~Kivelson et al., preprint

$\ \ \ $S.A.~Kivelson, M.I.~Salkola, Synth.Met.44, 281 (1991)

[7] B.C.~Xu and S.E.~Trullinger, Phys.Rev.Lett.57, 3113

[8] K.~Ziegler, Phys.Rev.Lett.59, 152 (1987), Z.Phys.B78, 281 (1990)

[9] D.~Boyanovsky, Phys.Rev.B 27, 6763 (1983)

[10] N.~Basescu et al., Nature 327, 403 (1987)

[11] K.G.~Wilson and J.~Kogut, Phys.Rep.C2, 75 (1974)
\vfill
\eject
\no
{\bf Figure Captions}
\bigskip
\no
Fig.1: Renormalization group flow of the spin-dependent interaction
$\g$ and the new interaction $\nu$. The initial value of the
fermion-phonon induced interaction is $\l_0=0.1$ and spin is $n=3$.
\bigskip
\no
Fig.2: Renormalization group flow of the fermion-phonon induced
interaction $\l$ and the spin-dependent interaction $\g$ for spin $n=3$.
\bigskip
\bye